\documentclass[fleqn,10pt]{wlscirep}
\title{Exchange-torque-induced excitation of perpendicular standing spin waves in nanometer-thick YIG films}

\author[1,*]{Huajun Qin}
\author[1]{Sampo J. H\"am\"al\"ainen}
\author[1,*]{Sebastiaan van Dijken}
\affil[1]{NanoSpin, Department of Applied Physics, Aalto University School of Science, P.O. Box15100, FI-00076 Aalto, Finland}

\affil[*]{huajun.qin@aalto.fi, sebastiaan.van.dijken@aalto.fi}

\begin{abstract}
Spin waves in ferrimagnetic yttrium iron garnet (YIG) films with ultralow magnetic damping are relevant for magnon-based spintronics and low-power wave-like computing. The excitation frequency of spin waves in YIG is rather low in weak external magnetic fields because of its small saturation magnetization, which limits the potential of YIG films for high-frequency applications. Here, we demonstrate how exchange-coupling to a CoFeB film enables efficient excitation of high-frequency perpendicular standing spin waves (PSSWs) in nanometer-thick (80 nm  and 295 nm) YIG films using uniform microwave magnetic fields. In the 295-nm-thick YIG film, we measure intense PSSW modes up to 10th order. Strong hybridization between the PSSW modes and the ferromagnetic resonance mode of CoFeB leads to characteristic anti-crossing behavior in broadband spin-wave spectra. A dynamic exchange torque at the YIG/CoFeB interface explains the excitation of PSSWs. The localized torque originates from exchange coupling between two dissimilar magnetization precessions in the YIG and CoFeB layers. As a consequence, spin waves are emitted from the YIG/CoFeB interface and PSSWs form when their wave vector matches the perpendicular confinement condition. PSSWs are not excited when the exchange coupling between YIG and CoFeB is suppressed by a Ta spacer layer. Micromagnetic simulations confirm the exchange-torque mechanism.  
\end{abstract}
\begin{document}

\flushbottom
\maketitle
% * <john.hammersley@gmail.com> 2015-02-09T12:07:31.197Z:
%
%  Click the title above to edit the author information and abstract
%
\thispagestyle{empty}

\section*{Introduction}

Magnonics aims at the use of spin waves for the processing, storage, and transmission of information \cite{Kruglyak2010, Serga2010, Khitun2010, Krawczyk2014, Chumak2015, Nikitov2015}. With the smallest damping parameter of all magnetic materials, ferrimagnetic YIG has attracted considerable interest. Several building blocks for spin-wave-based technologies have been realized using YIG magnonics, including magnonic crystals \cite{Chumak2008, Chumak2010, Nikitov2015}, logic gates \cite{Schneider2008}, transistors \cite{Chumak2014}, and multiplexers \cite{Davies2015}. In these experiments, coplanar waveguides (CPWs) or microstrip antennas are typically used to excite propagating magnetostatic spin waves. The frequency of these spin waves depends on their wave vector ($k$), the saturation magnetization of YIG ($M_{s}$), and the external magnetic field ($H_{ext}$). Because the saturation magnetization of YIG is small and the wave vector is limited by the width of the antenna signal line, the spin-wave frequency is only $1-2$ GHz in weak magnetic fields \cite{Yu2014, Collet2017}. Higher frequencies can be attained by the excitation of magnetostatic spin wave modes with larger wave vector using a grating coupler \cite{Yu2016}, at the expense of emission efficiency.      

Another spin-wave mode that can be excited in a magnetic film is the PSSW. The wave vector of this confined mode is approximated by $k = p\pi/d$, where $d$ is the film thickness and $p$ is the order number. Thus in nanometer-thick magnetic films, the wave vectors of PSSW modes are large and their frequency is high. The formation of a PSSW requires a nonuniform excitation across the magnetic film thickness. Laser pulses \cite{Demokritov2001, Busse2015, Razdolski2017}, microwave magnetic fields from a miniaturized antenna \cite{Klingler2015}, and eddy-current shielding in conducting films \cite{Kennewell2010} have been used to excite PSSWs. In these experiments, the excitation field is nonuniform and both odd and even PSSW modes are measured. Uniform microwave magnetic fields can also excite PSSWs if the magnetization of the film is pinned at one or both of its interfaces \cite{Kittel1958, Soohoo1963, Wigen1962, Gui2007, Khivintsev2010, Magaraggia2011, Schoen2015}. Symmetrical pinning only induces odd PSSW modes, whereas both odd and even modes should be detected if the magnetization is pinned at one of the interfaces. Most work on PSSWs has focussed on metallic ferromagnetic materials such as Co/Py \cite{Kennewell2010}, Py \cite{Gui2007, Khivintsev2010, Magaraggia2011, Schoen2015}, and CoFeB \cite{Conca2014,Navabi2017}. In YIG, Klingler \textit{et al.} used PSSWs to extract the exchange constant \cite{Klingler2015} and Navabi \textit{et al.} demonstrated the excitation of a 1st order PSSW mode in a 100-nm-thick YIG film on top of an undulating substrate \cite{Navabi2017}.          

Here, we report on efficient excitation of PSSWs in nanometer-thick YIG films. The excitation mechanism is based on exchange coupling between the YIG film and a CoFeB layer. We show that forced magnetization precessions in YIG and CoFeB, driven by an approximately uniform CPW field, induce a dynamic exchange torque at the interface when the precessions are dissimilar. Consequently, the emission of spin waves into YIG is most efficient if the dynamic exchange torque is maximized near the ferromagnetic resonance (FMR) frequency of either YIG or CoFeB. Because the PSSW dispersion relations cross the FMR curve of the CoFeB layer, PSSWs with high order numbers are efficiently excited in YIG at high frequencies. 

\section*{Results}

\subsection*{Spin-wave spectra of YIG, CoFeB, and YIG/CoFeB films}

\begin{figure}[ht]
\centering
\includegraphics[width=\linewidth]{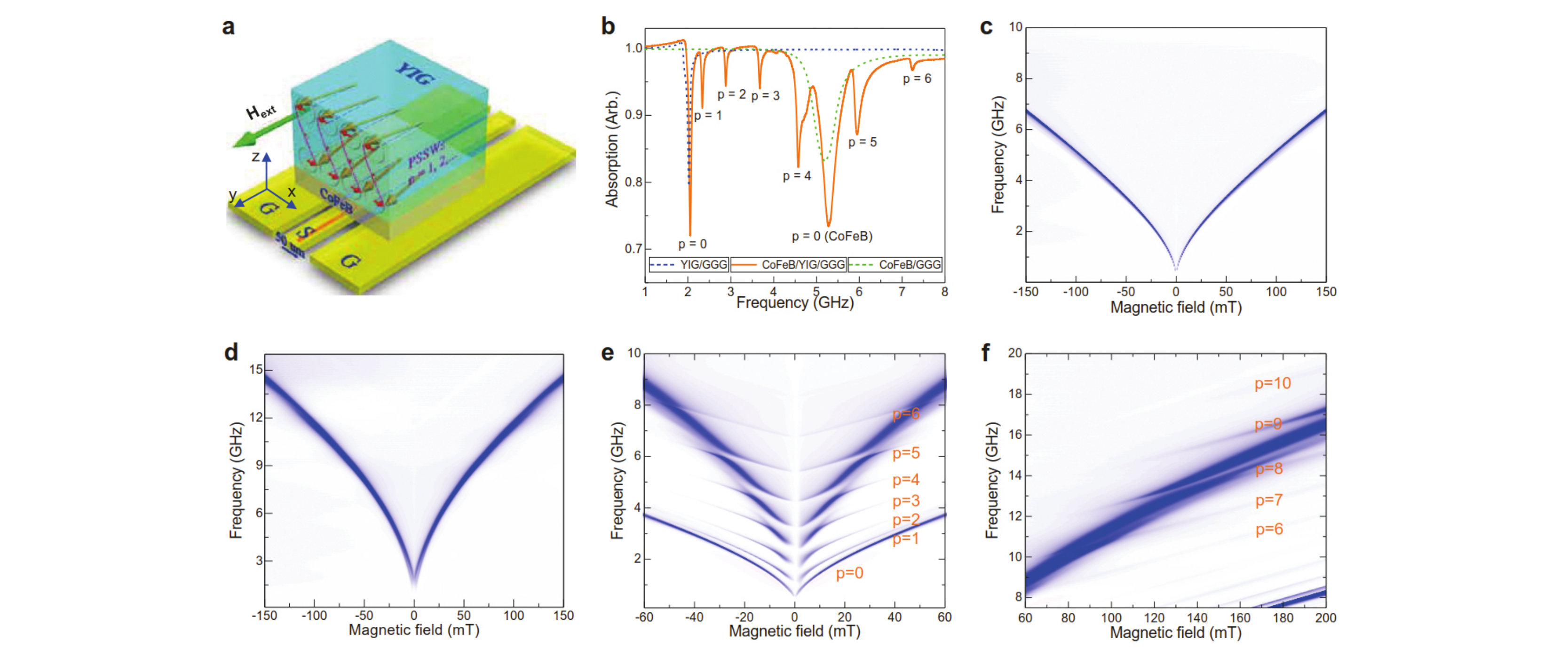}
\caption{(a) Schematic of the measurement geometry (not to scale). Spin-wave spectra are obtained by placing the sample face-down on a CPW. A microwave current is injected into the CPW signal line using a VNA. This produces a nearly uniform spin-wave excitation field. Spin-wave absorption is measured in transmission using scattering parameter $S_{12}$. The external magnetic bias field is oriented parallel to the CPW. (b) Spin-wave spectra for a single YIG film (dashed blue line), a single CoFeB film (dashed green line), and a YIG/CoFeB bilayer (solid orange line), measured with an external magnetic bias field of 20 mT. The YIG and CoFeB films are 295 nm and 50 nm thick. The FMR modes in YIG and CoFeB ($p$ = 0) and higher order PSSW modes in YIG ($p$ = 1 ... 6) are labeled. (c)-(e) Spin-wave spectra of the same samples as a function of magnetic bias field: (c) YIG, (d) CoFeB, (e) YIG/CoFeB. (f) YIG/CoFeB spin-wave spectrum at higher frequency and larger magnetic bias field, demonstrating the excitation of PSSWs with large order numbers.}
\label{Fig1}
\end{figure}

Single-crystal ferrimagnetic YIG films with a thickness of 80 nm and 295 nm were grown on (111)-oriented Gd$_3$Ga$_5$O$_{12}$ (GGG) substrates using pulsed laser deposition (PLD). To measure broadband spin-wave spectra, we placed the films face-down onto a CPW with a 50 $\mu$m-wide signal line. A microwave current provided by a vector network analyzer (VNA) was used to generate a microwave magnetic field around the CPW. The main excitation strength of the CPW was at wave vector $k\approx0$. We recorded absorption spectra in transmission by measuring the real part of the $S_{12}$ scattering parameter. The experiments were performed with an external magnetic bias along the CPW. A schematic of the measurement geometry is shown in Fig. \ref{Fig1}a. 

We first discuss the spin-wave spectrum of a single 295-nm-thick YIG film. Figure \ref{Fig1}b shows the absorption at a magnetic bias field of 20 mT (dashed blue line). Data as a function of magnetic field are shown in Fig. \ref{Fig1}c. Obviously, only one spin-wave mode is excited in the film. The mode corresponds to uniform magnetization precession in YIG, i.e., the FMR mode. Higher-order PSSW modes are not detected in this sample, as expected from the near-uniform microwave magnetic field and the absence of magnetization pinning at the interfaces. After characterization, we deposited a 50-nm-thick CoFeB layer onto the same YIG film using magnetron sputtering. Figure \ref {Fig1}b (solid orange line) and Fig. \ref {Fig1}e show the spin-wave spectrum of this sample. Now, a large number of spin-wave modes are measured. The lowest-frequency mode, labeled as $p$ = 0, corresponds to FMR in YIG (compare the blue and orange lines in Fig. \ref {Fig1}b or the spectra in Figs. \ref {Fig1}c and \ref {Fig1}e). The higher-order modes ($p$ = 1, 2, ...) are PSSWs in YIG (see next Section for details). The excitation of both odd and even modes implies that exchange coupling at the YIG/CoFeB interface produces an asymmetric pinning configuration. Finally, we note that a FMR mode is also excited in CoFeB ($p$ = 0 (CoFeB)). To support this conclusion, we show the spin-wave spectrum of a single 50-nm-thick CoFeB film on GGG in Fig. \ref {Fig1}b (dashed green line) and in Fig. \ref {Fig1}d.     

The data in Fig. \ref {Fig1} clearly indicate that the excitation of PSSWs in YIG is particularly efficient near the FMR of the CoFeB film. This point is further exemplified by the spin-wave spectrum of Fig. \ref {Fig1}f, demonstrating the formation of PSSW modes with order numbers up to $p$ = 10 at high frequencies. High-order PSSWs are only visible if their frequency approaches that of the $p$ = 0 mode in CoFeB. Exchange coupling at the YIG/CoFeB interface results in mode hybridization and characteristic anti-crossing behavior. Consequently, the frequency of the CoFeB FMR mode in the YIG/CoFeB bilayer is slightly shifted with respect to the same resonance in the single CoFeB film (orange and green curves in Fig. \ref {Fig1}b). 

In the following, we first analyze the PSSW modes in YIG/CoFeB using a phenomenological model. Next, we assess their intensity and linewidth. Results for YIG/CoFeB bilayers with a 80-nm-thick YIG film are subsequently discussed. Finally, we elucidate the origin of efficient PSSW excitation in exchange-coupled YIG/CoFeB bilayers using control experiments with a nonmagnetic spacer layer and micromagnetic simulations.  

\subsection*{Analysis of PSSW mode dispersions}
 
To investigate the dependence of the PSSW resonance frequency on external bias field, we first extract experimental data for $p$ = 0 ... 6 from the spin-wave spectra in Fig. \ref{Fig1}. The results are plotted as symbols in Fig. \ref{Fig2}a. We derive the saturation magnetization of our YIG film by fitting the frequency dependence of the $p$ = 0 mode to the Kittel formula \cite{Kittel1948}: $f = \gamma \mu_0/ 2\pi \sqrt {H_{ext} (H_{ext} + M_s)}$. Using $\gamma/ 2\pi$ = 28 GHz/T, we obtain a good fit for $M_s= 192$ kA/m. This magnetization value compares well to previous results on YIG films \cite {Howe2015, Sokolov2016}. Next, we fit the PSSW modes ($p$ = 1 ... 6) to the following dispersion relation \cite {Gui2007, Conca2014, Navabi2017}:
\\
\begin{equation} \label{Eq:PSSW}
f_{PSSW} = \frac{\gamma \mu_0} {2\pi} \sqrt{\bigg[H_{ext} + \frac{2A_{ex}}{\mu_0 M_{s}}\Big(\frac{p \pi}{d}\Big)^2\bigg]\bigg[H_{ext} + \frac{2A_{ex}}{\mu_0 M_{s}}\Big(\frac{p \pi}{d}\Big)^2+ M_{s}\bigg]},
\end{equation} 
\\
where $A_{ex}$ is the exchange constant. By inserting $\gamma/2\pi$ = 28 GHz/T, $M_{s}$ = 192 kA/m, $d$ = 295 nm, and $p=1-6$, we obtain good fits to all PSSW modes using $A_{ex}=3.1$ pJ/m (lines in Fig. \ref{Fig2}a). This value also agrees with literature \cite{Klingler2015}.

The agreement between our experimental data and the model confirms that the higher-order resonances in the spin-wave spectra of the YIG/CoFeB bilayer correspond to PSSW modes in YIG. The observation that integer numbers of $p$ and the actual thickness of the YIG film in Eq. \ref{Eq:PSSW} provide excellent fits to all dispersion curves over a large bias-field range signifies strong magnetization pinning at the YIG/CoFeB interface. Weak pinning at the boundary of the YIG film would require the use of $p-\Delta{p}$ in the fitting formula \cite {Gui2007}, where correction factors $\Delta{p}$ = 0 and $\Delta{p}$ = 0 would correspond to full and zero magnetization pinning, respectively. In our YIG/CoFeB bilayer, short-range exchange coupling between the two magnetic films provides a strongly pinned interface.       

\subsection*{Intensity and linewidth of PSSW resonances}

\begin{figure}[ht]
\centering
\includegraphics[width=0.5\linewidth]{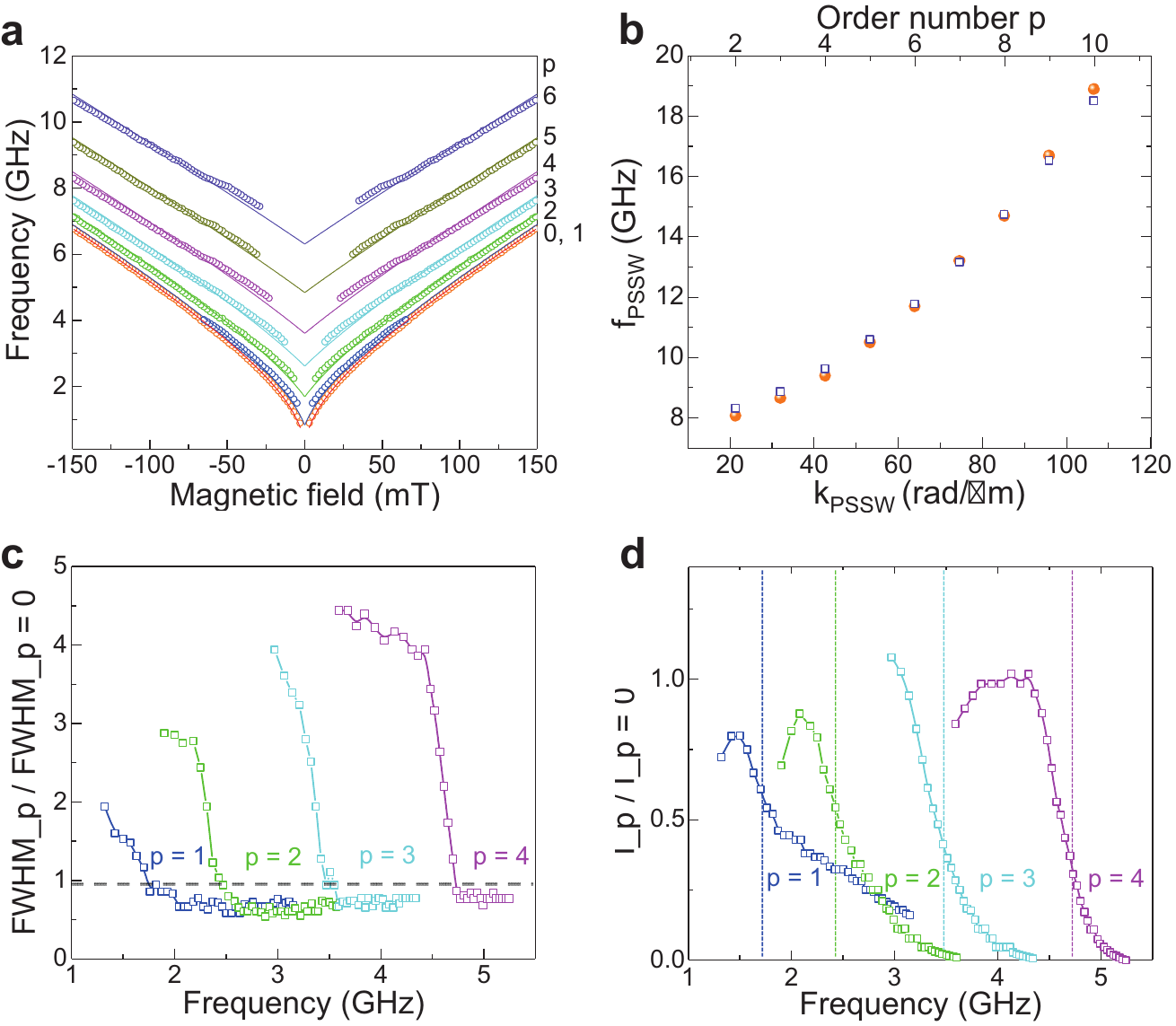}
\caption{(a) Frequency of the FMR and PSSW modes in a 295-nm-thick YIG film as a function of magnetic bias field. The symbols are extracted from the  experimental spin-wave spectrum in Fig. \ref{Fig1}e. The lines are fits to the data using the Kittel formula ($p$ = 0) and the PSSW dispersion relation (Eq. \ref{Eq:PSSW}, $p$ = 1 ... 6). (b) PSSW mode frequency as a function of wave vector for a magnetic bias field of 185 mT. The solid and empty symbols denote experimental data and calculated values (Eq. \ref{Eq:PSSW}), respectively. (c),(d) FWHM linewidth and intensity of the PSSW resonances as a function of frequency. The properties of the $p$ = 1 ... 4 modes are normalized to those of the FMR mode ($p$ = 0).}
\label{Fig2}
\end{figure}

Damping of PSSW modes in magnetic films  can have different origins. Besides intrinsic damping, eddy-current damping (in metallic films), and radiative damping caused by inductive coupling between the sample and the microwave antenna can also contribute \cite {Schoen2015}. To assess the damping of PSSWs in our YIG/CoFeB bilayer, we plot the full width at half maximum (FWHM) linewidth of the $p$ = 1 ... 4 modes relative to that of the $p$ = 0 mode (Fig. \ref{Fig2}c). The frequency evolution of this data was obtained from the spin-wave spectra in Fig. \ref {Fig1}e for magnetic fields ranging from 0 to 30 mT. The linewidths of all PSSW modes are large at low frequencies. The broad resonances in YIG are caused by hybridization with the higher-loss FMR mode in CoFeB. As the frequency increases, the frequency gap between the PSSWs in YIG and the FMR mode in CoFeB becomes larger. Once the two modes decouple, the linewidths of the PSSW modes decrease. In the decoupled state, the PSSW linewidths are similar to that of the $p$ = 0 mode in YIG, independent of frequency. Since eddy-current damping can be omitted in insulating YIG and radiative damping would cause the linewidth to increase with frequency \cite{Schoen2015}, the data in Fig. \ref{Fig2}c suggest that damping of PSSWs is dominated by intrinsic material parameters.    
	
Figure \ref{Fig2}d shows the relative intensity of the same PSSW modes. The dashed lines indicate the frequency where FWHM$_{p}$/FWHM$_{p = 0}$ = 1 in Fig. \ref{Fig2}c, which we use as an indicator for dehybridization between the PSSWs in YIG and the FMR mode in CoFeB. For the pure PSSW modes beyond this critical frequency, we still measure high intensities. The intensities of the $p$ = 1 and $p$ = 2 modes are up to 50\% of the $p$ = 0 resonance and this value drops to about 30\% for $p$ = 3 and $p$ = 4. The large intensities of the PSSW modes demonstrate a highly efficient excitation mechanism. 

\subsection*{Tuning of PSSW modes in YIG/CoFeB bilayers}

\begin{figure}[ht]
\centering
\includegraphics[width=0.5\linewidth]{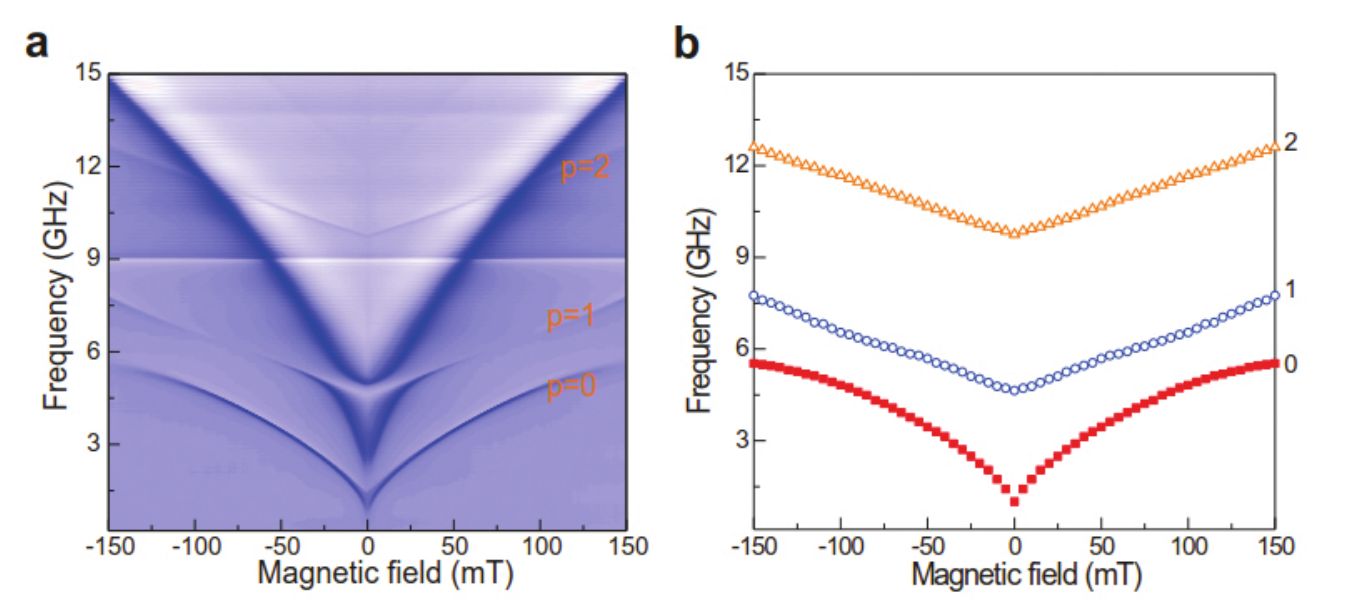}
\caption{(a) Spin-wave spectrum of a 80 nm YIG/50 nm CoFeB bilayer. (b) Extracted frequency of the $p$ = 0 ... 2 modes in YIG, demonstrating an up-shift in PSSW frequency compared to data for the 295-nm-thick YIG film (Fig. \ref{Fig2}a).}
\label{Fig3}
\end{figure}

The frequency of a PSSW depends on the wave vector of the confined mode and the external magnetic bias field. Since $k= p\pi/d$, the frequency of a PSSW could be enhanced by a reduction of the film thickness $d$. For an efficient excitation method, this would enable high-frequency spin waves in YIG at small magnetic fields. To test this prospect, we prepared a 80 nm YIG/50 nm CoFeB bilayer. The spin-wave spectrum of this sample is shown in Fig. \ref{Fig3}. In addition to the FMR modes in YIG and CoFeB, the first two PSSW modes are also measured. Anti-crossing behavior between the $p$ = 1 mode and the CoFeB resonance and an increase of the PSSW intensity near the anti-crossing frequency are again apparent. Compared to the 295-nm-thick YIG film, the PSSWs in thinner YIG are shifted up in frequency. At a moderate magnetic bias field of 20 mT, the increase of frequency amounts to about 3 GHz for $p$ = 1 and 8 GHz for $p$ = 2. The data of Fig. \ref{Fig3} thus confirm that PSSW modes are efficiently excited at high frequencies if the thickness of YIG is reduced.     

\subsection*{PSSW excitation mechanism}

We explain the excitation of PSSWs in YIG/CoFeB bilayers by a dynamic exchange torque at the interface. The uniform microwave excitation field from the CPW induces forced magnetization precessions in both magnetic layers. If the amplitudes of these precessions are different, a dynamic exchange torque is generated, causing the emission of spin waves from the interface. The efficiency of this excitation mechanism depends on the strength of the dynamic exchange torque, which is maximized at the FMR frequency of YIG and CoFeB. While spin waves are emitted from the interface over a broad frequency range, PSSWs only form when the wave vector of the excited spin waves matches the perpendicular confinement condition ($k = p\pi/d$) of the YIG film. Our experiments support this scenario. PSSWs are only measured after the YIG film is covered by a CoFeB layer and the PSSW resonances are most intense if the induced precession of magnetization is large in one of the two layers, i.e., near the FMR of YIG or CoFeB (see Figs. \ref{Fig1}b,e,f and Fig. \ref{Fig3}a). 

\begin{figure}[ht]
\centering
\includegraphics[width=0.5\linewidth]{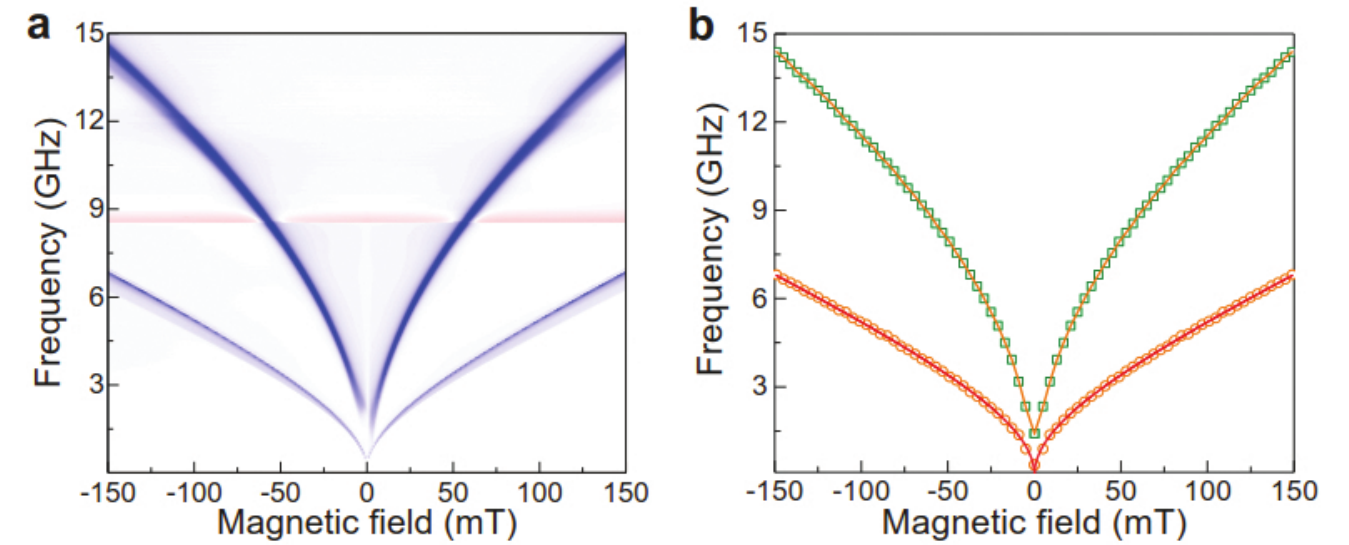}
\caption{(a),(b) Spin-wave spectrum for a 295-nm-thick YIG film and a 50-nm-thick CoFeB layer, separated by 10 nm of Ta. Only the FMR modes in YIG and CoFeB are measured in this case. The lines in (b) are fits to the two resonances using the Kittel formula.}
\label{Fig4}
\end{figure}

To confirm the crucial role of exchange coupling at the YIG/CoFeB interface, we prepared a 295 nm YIG/10 nm Ta/50 nm CoFeB trilayer on GGG. The spin-wave spectrum of this sample is shown in Fig. \ref{Fig4}. As expected, no PSSW modes are measured in this case. The two resonances in the spectrum are identical to those in Figs. \ref{Fig1}c and \ref{Fig1}d and, thus, correspond to the FMR mode in the YIG and CoFeB film, respectively. The Kittel formula fits the experimental data for $M_s$ = 192 kA/m (YIG) and $M_s$ = 1280 kA/m (CoFeB). The results of Fig. \ref{Fig4} demonstrate that the elimination of exchange coupling between magnetization precessions in YIG and CoFeB by the Ta spacer layer destroys the driving force behind PSSW excitation. This also implies that dipolar coupling between YIG and CoFeB is insignificant.    

\begin{figure}[ht]
\centering
\includegraphics[width=0.5\linewidth]{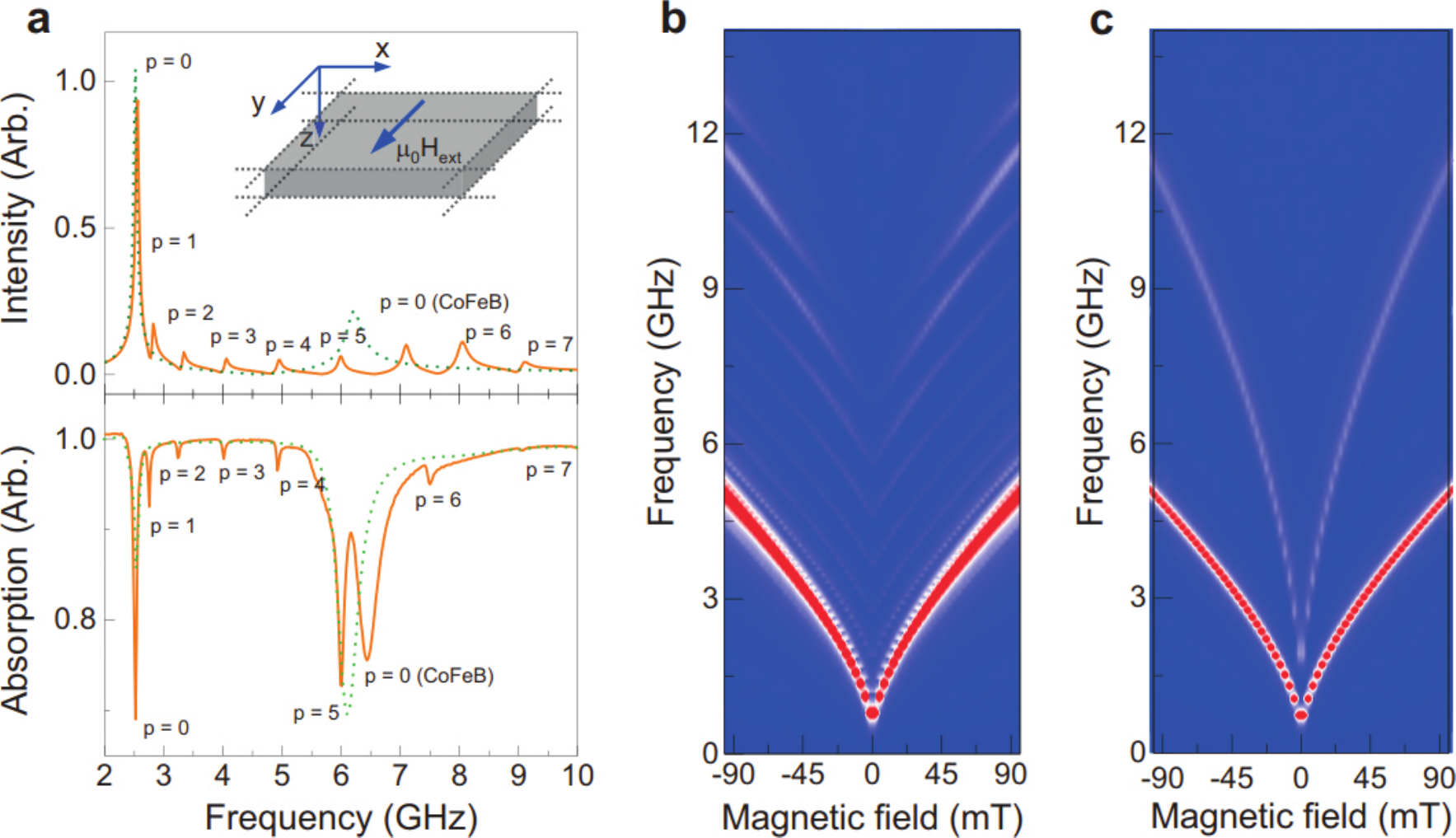}
\caption{(a) Simulated (top) and measured (bottom) spin-wave spectra for a 295 nm YIG/50 nm CoFeB bilayer (solid orange line) and a 295 nm YIG/10 nm Ta/50 nm CoFeB trilayer (dashed green line). The magnetic bias field is 30 mT. The inset illustrates the simulation geometry. (b),(c) Simulated spectra for the same structures as a function of magnetic bias field.}
\label{Fig5}
\end{figure}

We performed micromagnetic simulations in MuMax3 \cite{Vansteenkiste2014} to further study the microscopic origin of PSSWs in YIG/CoFeB bilayers. In the simulations, we considered a 295-nm-thick YIG film and a 50-nm-thick CoFeB layer. The structure was discretized using finite-difference cells of size $x$ = 54 nm, $y$ = 54 nm and $z$ = 2.7 nm, as schematically shown in the inset of Fig. \ref{Fig5}a. Two-dimensional periodic boundary conditions were applied in the film plane to mimic an infinite bilayer. We used the following input parameters: $M_{s}$ = 192 kA/m (YIG), $M_{s}$ = 1280 kA/m (CoFeB), $A_{ex}$ = 3.1 pJ/m (YIG), and $A_{ex}$ = 16 pJ/m (CoFeB). The damping constant was set to 0.005 for both magnetic films. For YIG, this relatively large value was selected to limit the computation time. Spin waves in the bilayer were excited by an uniform 3 mT sinc-function-type magnetic field pulse or a sinusoidal ac magnetic field (see Methods). The excitation field was along $x$ and a magnetic bias field was aligned along $y$. For comparison, we also performed micromagnetic simulations for a structure where the YIG and CoFeB films are separated by a 10-nm-thick nonmagnetic spacer, to mimic the response of a YIG/Ta/CoFeB trilayer.

The top panel of Fig. \ref{Fig5}a shows simulated spin-wave spectra for the YIG/CoFeB bilayer (solid orange line) and the YIG/Ta/CoFeB trilayer (dashed green line). The simulations were performed with a magnetic bias field of 30 mT. For comparison, we plotted the measured spectra of these samples in the bottom panel of Fig. \ref{Fig5}a. The simulations reproduce the excitation of PSSW modes in the YIG film of the YIG/CoFeB bilayer ($p$ = 1 ... 7) and the absence of these modes in YIG/Ta/CoFeB. The simulated PSSW frequencies are in good agreement with the experiments, except for frequencies near the CoFeB FMR mode. This discrepancy is attributed to stronger hybridization between the PSSWs and the CoFeB FMR mode in the simulations, caused by stronger exchange coupling at a perfectly flat interface. Mode hybridization also shifts the FMR mode in CoFeB. Results for YIG/Ta/CoFeB confirm this view. For this decoupled structure, the frequency of spin-wave resonances are the same in the simulations and experiments (dashed green curves in Fig. \ref{Fig5}a). We note that different parameters are plotted in the simulated and measured spectra. In the simulations, the intensity of the resonances is proportional to the amplitude of magnetization precession. The intensity of the modes in the experiments, on the other hand, are determined by induction-related absorption of a microwave current in the CPW. For constant magnetization precession, the absorption signal would increase with frequency. As a result, the relative intensity of the CoFeB resonance at higher frequency is larger in the lower panel of  Fig. \ref{Fig5}a. Simulated spin-wave spectra as a function of magnetic bias field for both structures are shown in Figs. \ref{Fig5}b and \ref{Fig5}c. 

\begin{figure}[ht]
\centering
\includegraphics[width=\linewidth]{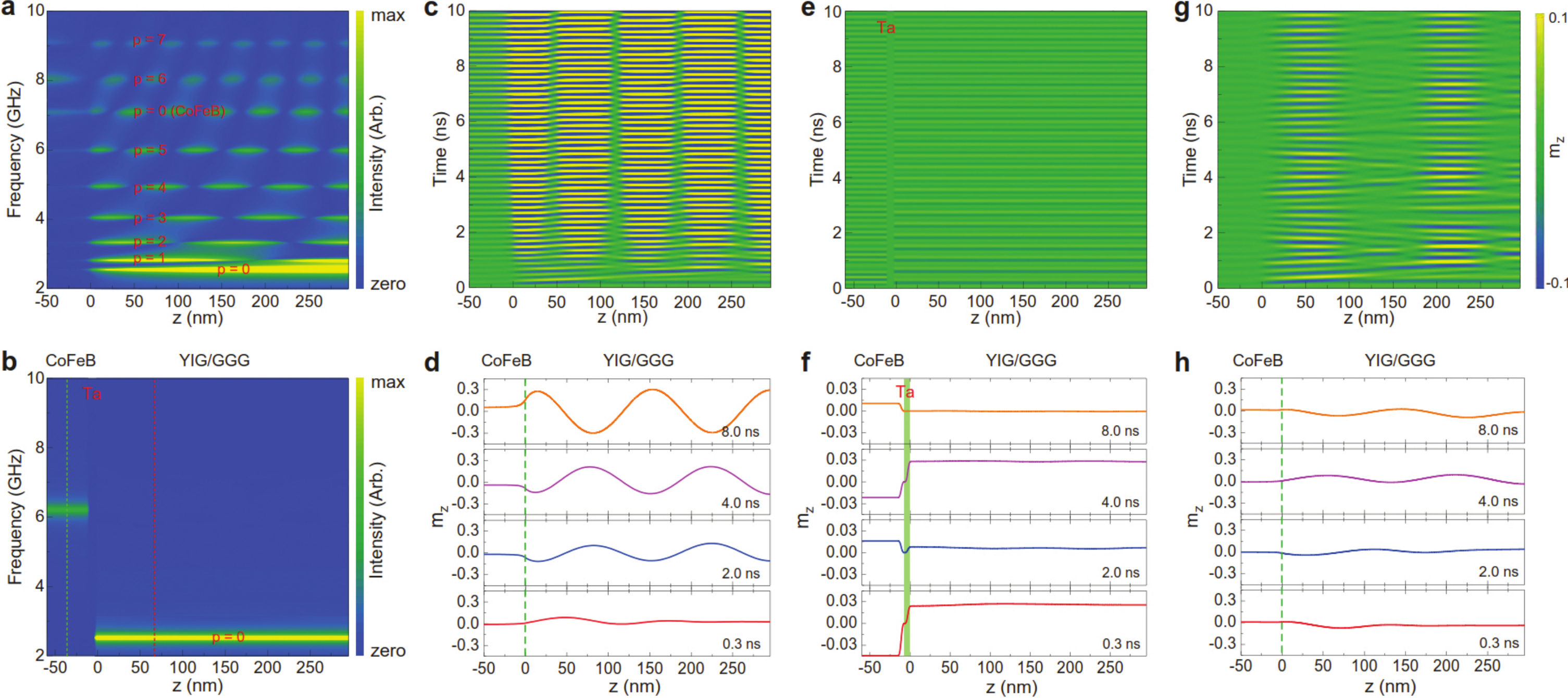}
\caption{(a),(b) Simulated spatial distribution of the FMR and PSSW modes in (a) a 295 nm YIG/50 nm CoFeB bilayer and (b) a 295 nm YIG/10 nm Ta/50 nm CoFeB trilayer. (c)-(f) Simulated time evolution of magnetization dynamics in (c),(d) a 295 nm YIG/50 nm CoFeB bilayer and (e),(f) a 295 nm YIG/10 nm Ta/50 nm CoFeB trilayer. The excitation frequency in these simulations is 4.9 GHz, which corresponds to the frequency of the $p$ = 4 PSSW mode. (g),(h) Simulated time evolution of magnetization dynamics in a 295 nm YIG/50 nm CoFeB bilayer at an off-resonance frequency of 4.5 GHz.}
\label{Fig6}
\end{figure}

We now focus on the spatial distribution of spin-wave modes in the YIG and CoFeB films. Figure \ref{Fig6}a and \ref{Fig6}b show simulation results for YIG/CoFeB and YIG/Ta/CoFeB, respectively. Magnetization precession in YIG at the FMR frequency is reduced near the CoFeB interface. This effect, which is absent in the YIG/Ta/CoFeB structure, signifies strong exchange coupling to the CoFeB layer. The PSSWs in YIG/CoFeB are nearly symmetric and strongly confined to the YIG film for the non-hybridized modes. The number of nodes corresponds to order parameter $p$. Hybridization between the PSSW modes in YIG and the FMR mode in CoFeB at higher frequencies also induces a significant magnetization precession in the CoFeB layer.    

The simulated time evolution of PSSW mode formation in the YIG/CoFeB bilayer is depicted in Figs. \ref{Fig6}c and \ref{Fig6}d. Here, we focus on $p$ = 4 at an excitation frequency of 4.9 GHz. The simulations illustrate how the magnetization responds to the onset of a spatially uniform sinusoidal ac magnetic field at $t$ = 0 s. Just after the excitation field is switched on, spin waves with a wavelength of $\lambda$ $\approx$ 150 nm ($\lambda=2d/p$) are emitted from the YIG/CoFeB interface. This excitation is triggered by a dynamic exchange torque originating from dissimilar magnetization precessions in the YIG and CoFeB layers. The emitted spin waves propagate along the thickness direction of the YIG film and reflect at the GGG/YIG interface. At the selected frequency of 4.9 GHz, the forward and backward propagating spin waves interfere constructively. As a result, a $p$ = 4 PSSW is formed. The large-amplitude PSSW is fully established after $t\approx$ 6 ns. 

The simulated time evolution of magnetization dynamics in the YIG/Ta/CoFeB trilayers is shown in Figs. \ref{Fig6}e and \ref{Fig6}f. As discussed previously, this structure does not support the excitation of PSSWs. Instead, the ac magnetic field induces uniform small-amplitude precessions of magnetization in the YIG and CoFeB layers. Because of different precession amplitudes, a time-dependent divergence of magnetization emerges at the location of the Ta insertion layer. This divergence is the source of the dynamic exchange torque in structures where the magnetization of YIG and CoFeB are directly coupled by interface exchange interactions.

Finally, we discuss the off-resonance time evolution of magnetization dynamics in the YIG/CoFeB bilayer (Figs. \ref{Fig6}g and \ref{Fig6}h). We consider an excitation frequency of 4.5 GHz, thus, in between the frequencies of the $p$ = 3 and $p$ = 4 PSSW modes (see Fig. \ref{Fig6}a). Under these circumstances, spin waves are again emitted from the YIG/CoFeB interface by the dynamic exchange torque. However, since the condition for constructive spin-wave interference along the film thickness of YIG is not fulfilled, their amplitude is not amplified and a PSSW does not form.    

In summary, we have demonstrated an efficient method for the excitation of PSSWs in nanometer-thick YIG films. The method relies on direct exchange coupling between the YIG film and a CoFeB top layer. The application of an uniform microwave magnetic field produces a strong dynamic exchange torque at the YIG/CoFeB interface. This results in short-wavelength spin-wave emission. A PSSW is excited if one of the perpendicular confinement conditions is met. Our findings open up a new route towards the excitation high-frequency spin waves in YIG. The results can also be generalized to other exchange-coupled systems. The excitation of intense PSSWs with large order numbers requires crossings between their dispersion relations and the FMR mode in a second, exchange-coupled magnetic layer. This situation is attained if the saturation magnetization is smallest in the PSSW carrying film. 

\section*{Methods}

\subsection*{Sample fabrication}

We grew YIG films with a thickness of 80 nm and 295 nm on single-crystal GGG(111) substrates using PLD. The GGG substrates were ultrasonically cleaned in acetone and isopropanol before loading into the PLD vacuum chamber. We degassed the substrates at $550^{\circ}$C for 15 minutes. After this, oxygen was inserted into the chamber. After setting the oxygen pressure to 0.13 mbar, we increased the temperature to 800 $^{\circ}$C at a 5 $^{\circ}$C per minute rate. The YIG films were deposited under these conditions from a stoichiometric target. We used an excimer laser with a pulse repetition rate of 2 Hz and a laser fluence of 1.8 J/cm$^{2}$. After film growth, we annealed the YIG films at $730^{\circ}$C in an oxygen environment of  13 mbar. The annealing time was 10 minutes. This was followed by a cool down to room temperature at a rate of $-3^{\circ}$C per minute. The deposition process resulted in single-crystal YIG films, as confirmed by X-ray diffraction. The composition of CoFeB was 40\%Co, 40\%Fe, and 20\%B. The CoFeB and Ta layers were grown by magnetron sputtering at room temperature.   

\subsection*{Spin-wave spectroscopy}

We recorded spin-wave absorption spectra in transmission by measuring the real part of the $S_{12}$ scattering parameter. To enhance contrast, a reference spectrum taken at larger magnetic field or frequency was subtracted from the measurement data. The setup consisted of a two-port VNA and a quadruple electromagnet probing station. The CPW with a 50 $\mu$m-wide signal line and two 800 $\mu$m-wide ground lines was patterned on a GaAs substrate. The gap between the signal and ground lines was 30 $\mu$m. The CPW was designed to provided a $k\approx0$ excitation field in the plane of the YIG film. During broadband spin-wave spectroscopy measurements, the sample was placed face-down onto the CPW. 

\subsection*{Micromagnetic simulations}

We performed micromagnetic simulations using open-source GPU-accelerated MuMax3 software. A $6900\times6900\times345$ nm$^3$ CoFeB/YIG bilayer structure was discretized into $54\times54\times2.7$ nm$^3$ cells and two-dimensional periodic boundary conditions were applied in the film plane. We abruptly changed the magnetic parameters at the YIG/CoFeB interface and used the harmonic mean value of the exchange constants in YIG and CoFeB to simulate the interface exchange coupling. The system was initialized by an external magnetic field along the $y$ axis followed by relaxation to the ground state. After this, a spatially uniform 3 mT sinc-function-type magnetic field pulse with a cut-off frequency of 20 GHz was applied along the $x$ axis. The magnetic field pulse excited all spin-wave modes up-to the cut-off frequency with uniform excitation power (Fig. \ref{Fig5}). To study the spatial dependence of magnetization dynamics (Fig. \ref{Fig6}), the system was driven be a sinusoidal ac magnetic field with an amplitude of 3 mT. In these simulations, the time evolution of the perpendicular magnetization component ($m_z$) was recorded for 50 ns in 3 ps time steps along the thickness direction of the system at the center of the simulation mesh. The spatially-resolved intensity was obtained by applying a Fourier imaging technique where the time evolution of $m_z$ was Fourier-transformed on a cell-by-cell basis. 

%\bibliography{./Qin}

\section*{Acknowledgements}

This work was supported by the European Research Council (Grant No. ERC-2012-StG 307502-E-CONTROL). S.J.H. acknowledges financial support from the V\"ais\"al\"a Foundation. Lithography was performed at the Micronova Nanofabrication Centre, supported by Aalto University.

\section*{Author contributions statement}

H.J.Q., S.J.H., and S.v.D. designed and initiated research. H.J.Q. fabricated the samples and conducted the measurements. H.J.Q. and S.J.H. performed the micromagnetic simulations. S.v.D. supervised the project. H.J.Q. and S.v.D. wrote the manuscript, with input from S.J.H. 

\section*{Additional information}

\textbf{Competing financial interests:} The authors declare that they have no competing interests. 

\end{document}